\documentclass[sigconf]{acmart}

\AtBeginDocument{%
  }

\setcopyright{acmlicensed}
\copyrightyear{2024}
\acmYear{2024}
\acmDOI{XXXXXXX.XXXXXXX}

\acmConference[VINCI 2024]{The 17th International Symposium on Visual Information Communication and Interaction}{December 11--13, 2024}{Hsinchu, Taiwan}
\acmBooktitle{The 17th International Symposium on Visual Information Communication and Interaction (VINCI 2024), December 11--13, 2024, Hsinchu, Taiwan}
\acmISBN{978-1-4503-XXXX-X/18/06}

\begin{document}

\title{Investigating Size Congruency Between the Visual Perception of a
VR Object and the Haptic Perception of Its Physical World Agent}


\author{Wenqi ZHENG}
\orcid{1234-5678-9012}
\affiliation{%
  \institution{The Hong Kong University of Science and Technology (Guangzhou)}
  \city{Guangzhou}
   \state{Guangdong}
   \country{China}
}
 \email{wzheng738@connect.hkust-gz.edu.cn}
 
\author{Dawei XIONG}
\orcid{1234-5678-9012}
\affiliation{%
  \institution{The Hong Kong University of Science and Technology (Guangzhou)}
   \city{Guangzhou}
   \state{Guangdong}
   \country{China}
}
\email{dxiong107@connect.hkust-gz.edu.cn}

\author{Cekai WENG}
\orcid{1234-5678-9012}
\affiliation{%
  \institution{The Hong Kong University of Science and Technology (Guangzhou)}
   \city{Guangzhou}
   \state{Guangdong}
   \country{China}
 }
\email{cweng368@connect.hkust-gz.edu.cn}

\author{Jiajun JIANG}
\orcid{1234-5678-9012}
\affiliation{%
  \institution{The Hong Kong University of Science and Technology (Guangzhou)}
   \city{Guangzhou}
   \state{Guangdong}
   \country{China}
}
\email{jjiang127@connect.hkust-gz.edu.cn}

\author{Junwei Li}
\affiliation{%
  \institution{The Hong Kong University of Science and Technology (Guangzhou)}
   \city{Guangzhou}
   \state{Guangdong}
   \country{China}
 }
 \email{jli801@connect.hkust-gz.edu.cn}
\orcid{1234-5678-9012}

 \author{Jinni Zhou}
\orcid{1234-5678-9012}
\affiliation{%
  \institution{The Hong Kong University of Science and Technology (Guangzhou)}
   \city{Guangzhou}
   \state{Guangdong}
   \country{China}
 }
\email{eejinni@hkust-gz.edu.cn}
 
 \author{Mingming Fan}
\authornotemark[1]
\orcid{1234-5678-9012}
\affiliation{%
  \institution{The Hong Kong University of Science and Technology (Guangzhou)}
    \city{Guangzhou}
   \state{Guangdong}
   \country{China}
    \institution{ Hong Kong University of Science and Technology}
 }
 \email{mingmingfan@ust.hk}
 
\renewcommand{\shortauthors}{Zheng et al.}

\begin{abstract}
  Sandplay is an effective psychotherapy for mental retreatment, and many people prefer to engage in sandplay in Virtual Reality (VR) due to its convenience. Haptic perception of physical objects and miniatures enhances the realism and immersion in VR. Previous studies have rendered sizes by exerting pressure on the user’s fingertips or employing tangible, shape-changing devices. However, these interfaces are limited by the physical shapes they can assume, making it difficult to simulate objects that grow larger or smaller than the interface.
Motivated by literature on visual-haptic illusions, this work aims to convey the haptic sensation of a virtual object’s shape to the user by exploring the relationships between the haptic feedback from real objects and their visual renderings in VR. Our study focuses on the confirmation and adjustment ratios for different virtual object sizes. The results show that the likelihood of users confirming the correct size of virtual cubes decreases as the object size increases, requiring more adjustments for larger objects. This research provides valuable insights into the relationships between haptic sensations and visual inputs, contributing to the understanding of visual-haptic illusions in VR environments.
\end{abstract}


\begin{teaserfigure}
 \centering
  \includegraphics[width=0.6\textwidth]{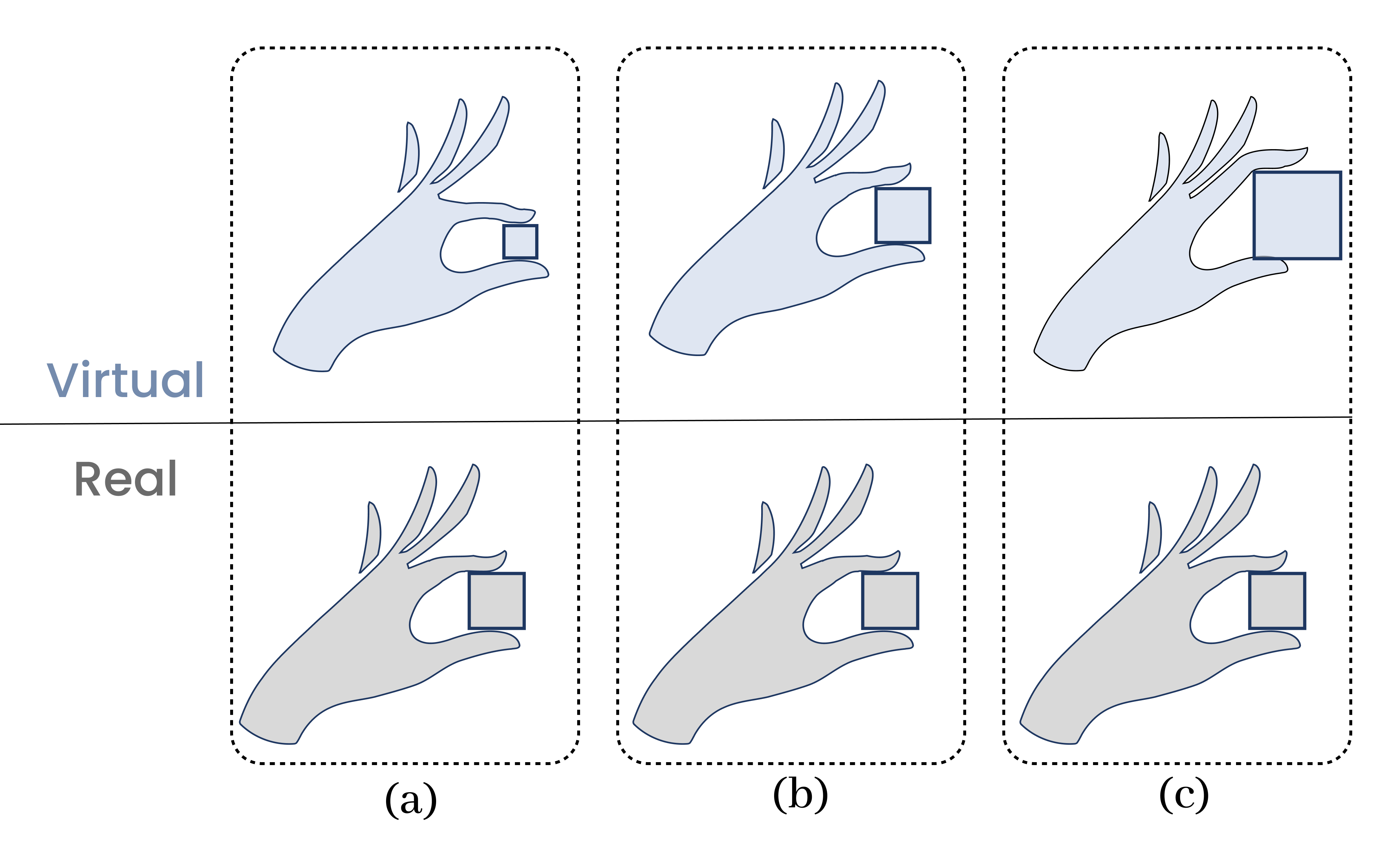}
  \caption{Study Overview: The user touches a block in reality while seeing their virtual hand touch a virtual block in VR and judges the size. (a): The user sees a block in VR that is smaller than the block they are touching in reality. (b): The user sees a block in VR that is the same size as the block they are touching in reality. (c): The user sees a block in VR that is larger than the block they are touching in reality.}
  \Description{Enjoying the baseball game from the third-base
  seats. Ichiro Suzuki preparing to bat.}
  \label{fig:teaser}
\end{teaserfigure}

\maketitle

\section{Introduction}
Sandplay therapy is a form of psychotherapy where clients create three-dimensional scenes in a tray filled with sand using various miniatures\cite{kalff2003sandplay}. With advancements in digital technology, digital methods, such as virtual reality (VR), are being explored to address the portability issues associated with traditional sandplay therapy, which involves physical sand trays and numerous miniatures. A simple approach is to render all miniatures visually in VR, enabling users to design and place their own items on the virtual tray\cite{hancock2010supporting}. However, the haptic feedback from physical miniatures is crucial for effective sandplay therapy\cite{wang2023ivrsandplay}. Different miniatures vary in size, and accurately rendering these size variations in a virtual environment can enhance user immersion in VR sandplay.

To simulate different sizes of virtual objects, previous research has introduced various haptic interfaces. Most of these interfaces utilize wearables\cite{choi2017grabity, choi2016wolverine, fang2020wireality} and exoskeletons\cite{bouzit2002rutgers, gu2016dexmo, jo2019evaluation} to restrict finger movements and simulate contact with virtual objects. However, each approach has limitations in replicating the dynamic resizing of virtual objects. Wearables and exoskeletons that restrict finger movement face specific challenges. Passive wearables, such as those using braking mechanisms, cannot effectively convey dynamically resizing objects\cite{choi2017grabity, choi2016wolverine, fang2020wireality, sinclair2019capstancrunch}. In contrast, active devices capable of simulating size changes tend to be heavy and power-intensive\cite{choi2018claw}. Another method involves using real objects to mimic the size of the virtual objects physically\cite{yoshida2020pocopo, gonzalez2021x, teng2018pupop}. While this approach can provide the best haptic feedback for VR users, it is limited by the impracticality of preparing objects in every possible size.

To address the aforementioned issues and provide the best immersive sandplay experience for users, we propose using real objects for size rendering. Our research aims to explore the relationships between the haptic feedback from real objects and their visual renderings in VR. A physically fixed-size object has the potential to be mapped to a range of virtual sizes through haptic sensing. This research provides insights into haptic-visual illusions, demonstrating how continuous visual changes can be effectively represented by discrete haptic inputs.


\section{Related work}

\subsection{Shape-changing displays}

 To achieve the haptic feedback of various virtual object forms, previous research has explored shape-changing displays. These shape displays\cite{siu2018shapeshift} can produce dynamic shapes and surfaces, typically using a planar array of linearly-actuated pins that move up and down, forming a 2.5D surface. To reduce the size of tabletop pin arrays, smaller handheld displays have been developed for shape rendering on the palm\cite{yao2013pneui} and index fingertip\cite{benko2016normaltouch}. Nonetheless, most shape displays remain bulky, complex, and limited in the area they can cover.
 
Various efforts have been made to create shape displays capable of rendering objects of different sizes in the hand. For instance, X-Rings\cite{gonzalez2021x} is a shape display designed to extrude surfaces 360 degrees around a central axis, allowing it to be grasped by the entire hand, although its ability to render dynamic shape changes is limited. Inflatable bladders \cite{bouzit2002rutgers, harrison2009providing, sareen2017printflatables, yao2013pneui} can shape the user’s hand by adjusting air pressure, producing a few fixed shapes and varying assistive forces. PuPoP\cite{teng2018pupop} can transition between a few fixed shapes using multiple inflatable bladders, while the commercial HaptX device covers the skin of the palm with tiny inflatable bubbles that simulate dynamic skin contact, although they cannot reproduce complete geometry. However, these approaches for dynamic real-time haptic sensations are limited by their reliance on additional pneumatic or hydraulic pumps and their restricted degrees of freedom.

\subsection{Finger haptic rendering}

Shape-changing displays have limitations in rendering the complete shape of a virtual object because parts of the object that do not come into contact with the user's skin are not sensed, potentially making it inefficient to render those parts. As an alternative, haptic exoskeletons \cite{bouzit2002rutgers, blake2009haptic, choi2016wolverine, fang2020wireality, in2011jointless, jo2019evaluation} provide feedback for rigid grasping through mechanical structures worn on the fingers. Most exoskeletons focus on rendering sensations at the fingertips. However, a significant drawback of haptic exoskeletons is their bulky design, making them cumbersome to wear.

Another approach employs handheld controllers that restrict finger movement to prevent them from closing on virtual objects. For example, NormalTouch \cite{benko2016normaltouch} uses a motorized platform that pushes the index finger back whenever it penetrates a virtual object. The CLAW controller \cite{choi2018claw} simulates a held object by applying force between the index finger and thumb, using a strong, heavy motor to resist user-applied forces. 

Wolverine\cite{choi2016wolverine} is a portable wearable haptic device designed to allow users to grasp rigid virtual objects. The authors created a light and low-cost device that renders force between the user's thumb and the three other fingers. To render the thickness and slipping of a virtual object pinched between two fingers, SpinOcchio\cite{kim2022spinocchio} uses two pivoting spinning disc modules, a set of width-change actuators, and a VIVE tracker for generating continuous skin-slip in 6-DoF and varying normal forces. Most of these types of methods require a designed device to render the feeling of grasping, which is inconvenient for the users due to the size and weight of the device. FingerX \cite{tsai2022fingerx} features an extending structure connected to each finger, rendering grounded penetration prevention forces when virtual objects are near physical surfaces like a table. Recently, an electro-tactile feedback system\cite{tanaka2023full} has been designed for full-hand touch rendering, using electrodes outside of the palmer. The method has excellent rendering results. However, the user still needs to wear the electrodes for haptic rendering.

\subsection{Visual-haptic illusion}
The challenge of rendering a wide range of shapes prompted us to seek a method to convey the haptic sensation of a virtual object's shape to the user. Traditional shape displays and haptic controllers often combine visual displays, which show the desired motion and shape, with limited haptic feedback to enhance the perception of the experience. For instance, Abtahi et al. \cite{abtahi2018visuo} used scaling and redirection of the user's hand to manipulate hand-eye coordination and extend the perceived resolution of a shape display. Gonzalez et al. \cite{gonzalez2021x} adapted the limited dynamic range of the X-Rings shape display to represent objects of larger scale and varying geometry. These approaches leverage the dominance of visual perception over haptic sensations, relying on the device's ability to effectively render the shape and provide the user with the experience of haptic feedback \cite{pusch2011pseudo}.

A conceptually similar work to this research is Kim et al\cite{kim2024big}. Unlike the previously mentioned studies, \cite{kim2024big} utilizes real objects for haptic input and renders visual-haptic feedback by adopting a finger-repositioning concept. The core idea is to determine the relationship between finger positions on the cylinder and the visually rendered size in VR. Similarly, Xiong \cite{xiong2024reach} exploited the fact that virtual reality (VR) can provide a realistic experience and explored the effects of different virtual hand redirection techniques on training motivation in rehabilitation users. Our work goes beyond finger-repositioning, focusing on a broader concept of visual-haptic illusions, aiming to uncover the universal relationships between the haptic sensation of size and visual input.

\section{Study 1: Investigating perceivable size differences for VR users}
In order to investigate the effect of size on the participant's perception of inconsistencies between objects in reality and in VR, we divided the study into two parts. In Study 1, we recruited six subjects to determine which sizes make the difference in size apparent to participants.
\subsection{Participants}
Six participants (3 females, 3 males, Table 1) took part in the study. All participants had both hands healthy and all had normal tactile sensation. To control for extraneous variables, we asked participants to report their dominant hand and asked them to use their dominant hand for the study. We measured the length between the top thumb and forefinger of a hand of the six participants, which were distributed between 11 and 14.2 cm.

\subsection{Materials}
Using 3D printing, we created square cubes with side lengths ranging from 1cm to 11cm, with a 1cm difference in side length between each cube. The color and material of each cube remains consistent. We created an experimental setup using a box in which the participant could put one hand into the box, but could not see what was going on in the box. Instead, the experimenter could observe what was going on inside the box from the other side and replace the cube that the participant was touching inside the box.

\begin{table}[t!]
\centering
\caption{Summary of Experiment Results (Units: cm)}
\label{tab:summary}
\begin{tabular}{p{1.8cm}p{1.8cm}p{1.8cm}p{1.8cm}}
\hline
\multicolumn{4}{c}{Demographic Information} \\ \hline
Participants & Gender & Dominant Hand & Length  \\ \hline
A1           & Male   & Right         & 13.5   \\
A2           & Female & Right         & 11.0   \\
A3           & Male   & Right         & 14.2   \\
A4           & Female & Right         & 12.1   \\
A5           & Male   & Left          & 13.5   \\
A6           & Female & Right         & 11.5   \\ \hline
\multicolumn{4}{c}{First Experiment} \\ \hline
Participant & Large Sizes  & Medium Sizes  & Small Sizes  \\ \hline
A1          & 9, 10, 11   & 6, 7, 8      & 1, 2, 3, 4, 5 \\
A2          & 9, 10, 11   & 6, 7, 8      & 1, 2, 3, 4, 5 \\
A3          & 9, 10, 11   & 6, 7, 8      & 1, 2, 3, 4, 5 \\
A4          & 8, 10, 11   & 5, 6, 7, 9   & 1, 2, 3, 4    \\
A5          & 8, 9, 10, 11& 6, 7         & 1, 2, 3, 4, 5 \\
A6          & 8, 9, 10, 11& 5, 6, 7      & 1, 2, 3, 4    \\ \hline
\multicolumn{4}{c}{Second Experiment} \\ \hline
Participant & Large Sizes  & Medium Sizes  & Small Sizes  \\ \hline
A1          & 9, 10, 11   & 5, 6, 7, 8   & 1, 2, 3, 4   \\
A2          & 8, 9, 10, 11& 6, 7         & 1, 2, 3, 4, 5\\
A3          & 8, 9, 10, 11& 6, 7         & 1, 2, 3, 4, 5\\
A4          & 9, 10, 11   & 5, 6, 7, 8   & 1, 2, 3, 4   \\
A5          & 7, 8, 9, 10,11& NaN        & 1, 2, 3, 4, 5, 6\\
A6          & 8, 9, 10, 11& 6, 7         & 1, 2, 3, 4, 5\\ \hline
\multicolumn{4}{c}{Third Experiment} \\ \hline
Participant & Large Sizes  & Medium Sizes  & Small Sizes \\ \hline
A1          & 9, 10, 11   & 7, 8         & 1, 2, 3, 4, 5, 6 \\
A2          & 8, 9, 10, 11& 6, 7         & 1, 2, 3, 4, 5    \\
A3          & 8, 9, 10, 11& 5, 6, 7      & 1, 2, 3, 4       \\
A4          & 8, 9, 10, 11& 5, 6, 7      & 1, 2, 3, 4       \\
A5          & 8, 9, 10, 11& 7            & 1, 2, 3, 4, 5, 6 \\
A6          & 8, 9, 10, 11& 5, 6, 7      & 1, 2, 3, 4       \\ \hline

\end{tabular}
\end{table}
\vspace{-10pt}

\subsection{Procedure}
To minimize selection bias due to participants' prior knowledge of the purpose of the experiment, we only told them to take the haptic feedback test. We asked participants to place their hands in the experimental setup and only allowed them to pinch the cubes with their index fingers and thumbs in order to perceive the size of the cube. First, we arranged the cubes in order from smallest to largest and handed them to the participants in order for them to touch them in the box. Afterward, we placed the cubes in random order and handed them to participants to touch in turn, and after completing each touch, participants were asked to report whether the size of the cube they had just touched was large, medium, or small. To ensure the stability of participants' perception of cube size, we conducted a total of three sets of experiments in which participants touched the cubes in a randomized order in each set of experiments.
Participants were informed in advance of the size of the cubes they would touch and were allowed to observe them without touching. In this way they would have a rough impression of the cube they would be touching, but there would be no tactile feedback as a prior knowledge to interfere with their subsequent judgment.
Through this method, we were able to understand the approximate range of participants' perceptions of the sizes of the cubes being touched.

\subsection{Data analysis}
We averaged the side lengths reported by the experimenter for each of the three types of cubes: large, medium, and small. We also conducted a regression test with thumb tip to index finger tip length as an influencing factor to exclude the effect of different hand sizes on the experimental results. We averaged the side lengths of cubes perceived by all participants for large, medium, and small sizes, respectively.
\subsection{Findings}
The results of our experiments show that the majority of participants regarded cubes with dimensions of 8cm x 8cm and larger as large sizes. Only one participant reported a 7cm x 7cm cube as a larger cube in one of the experiments. All participants reported the cubes smaller than 4cm x 4cm are small sizes, and a few participants reported the 5cm x 5cm and 6cm x 6cm cubes as small sizes. Almost all participants viewed 7cm x 7cm as a medium-sized cube, but some participants reported cubes between 5cm x 5cm and 8cm x 8cm as medium-sized cubes. We averaged the side lengths of cubes perceived by all participants for large, medium, and small sizes, respectively. We obtained results of 9.65 cm for the large cube side length, 6.57 cm for the medium cube, and 2.89 cm for the small cube (all retained to two decimal places). Our regression results showed that neither hand size had a significant effect on how participants perceived the square.
Therefore, in Study 2, we will use our derived large, medium, and small cube sizes, as standard sizes, to investigate the effect of size on perceived inconsistency.

\section{Study 2: the impact of haptic feedback on size judgments in VR}
\subsection{Proposed interaction concepts}
When people grasp objects, they can simultaneously gain an overall perception of the size of the object and its shape through a variety of cues, including tactile (grip sensation), kinesthetic (muscle tension), proprioceptive cues (hand width contrast), and visual cues.
For example, when people gently hold a thick dictionary and heavy textbook in their hand, the book completely fills their hand, providing tactile stimulation to the palm and fingers. Now maintaining fingertip contact, if people replace it with another thin magazin, they would notice that the block no longer fills the interior of the palm and has less contact with the skin surface, resulting in more empty space in the hand.
In the above example, we propose to pay attention to two main tactile cues (surface contact) and proprioceptive cues (finger posture), both of which give rise to the perception of size change. Thus, we hypothesize that one can perceive changes in physical object size to some extent through appropriate haptic cues, even when keeping the visual cues consistent.

In virtual environments, participants observe the size of virtual objects visually while interacting with virtual objects through actual hand touch to perceive their size changes. In addition, to help participants perceive size changes more accurately, we usedprogress bars or other interactive elements to resize objects in the virtual environment to match the participant's touch perception.

\subsection{Design and implementation}

We proposed a replacement size mechanism to replace the actual size held in the participant's hand and displayed in the VR in our study. The participant's seat height, viewpoint height, and table height in VR were consistent with reality. By providing contextually appropriate visuals, participants could interpret these cues as indicating changes in the size of the object in their hand. We provided the participant with trial rights and size manipulation bars.

\begin{itemize}
    \item \textbf{Unity program architecture}: In Unity, we implemented the following program structure to support the study: The Main Scene contains the participant's VR environment, including the virtual table and objects, and we set the participant's perspective to match the real environment. The Object Manager manages the size and position of virtual objects, resizing them according to the participant's actions. The Touch Sensing Module captures the participant's hand position and gestures, providing tactile feedback and interaction cues. The UI Controller includes size manipulation bars and confirmation buttons, displaying the current virtual object size and allowing the participant to make adjustments. Lastly, the Data Logging Module logs the participant's interaction data, including the size of each adjustment and the confirmed size, and allows for exploring the user study data.
\end{itemize}
\begin{itemize}
    \item \textbf{Implementation Details}: Our system uses the Text Mesh Pro
plugin to achieve accurate text display, ensuring that participants can
clearly see interaction cues in VR. It integrates the VR headset and
joystick through the Pico 4 SDK to realize high-precision hand
tracking and interaction. Additionally, scripting in Unity allows for
the dynamic resizing of virtual objects and real-time updates to the
UI display.
\end{itemize}

\begin{figure}[t!]
\centering
\includegraphics[width=1.0\linewidth]{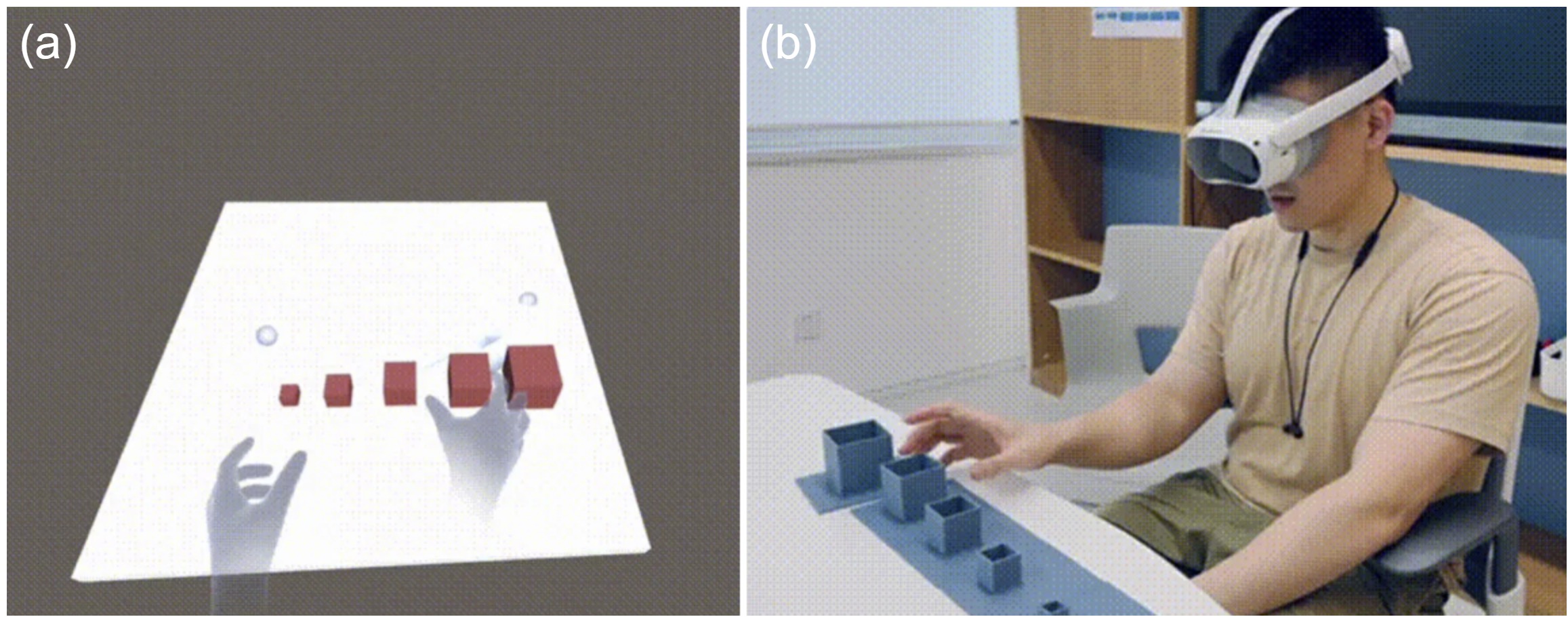}
\caption{Experiment environment. 
(a) shows the experimental setup in VR. Participants face a white table with a cube on it. Under the guidance of the experimenter, participants need to touch the cube with a virtual hand and report whether the size matches the real cube they have touched. The other cubes in the figure represent the different sizes we will ask participants to touch, but only one cube will be displayed during the experiment.
(b) shows the experimental setup in the real world. While participants touch the cube with a virtual hand, the experimenter will guide them to touch a real cube.}

\label{fig:imple}
\end{figure}

\subsubsection{Implementation Details}

Our system uses the Text Mesh Pro plugin to achieve accurate text display, ensuring that participants can clearly see interaction cues in VR. It integrates the VR headset and joystick through the Pico 4 SDK to realize high-precision hand tracking and interaction. Additionally, scripting in Unity allows for the dynamic resizing of virtual objects and real-time updates to the UI display.

\subsection{Experimental design}
The purpose of this study was to validate the participant's ability to perceive changes in the size of an object through haptic cues in a virtual environment. Participants would hold three sizes of physical cubes (small, medium, and large) and adjust the size of the virtual cubes in VR to match the held physical cubes.

The steps of the experiment are as follows:

\begin{itemize}
    \item \textbf{Size judgment using finger repositioning (i.e., pinching fingers) alone}: participants first perceived the size of the virtual cube by pinching their fingers. If they perceived that the size of the virtual cube matched that of the real one, they confirmed it; if it did not, only then did they proceed to the next step of adjustment.
\end{itemize}

\begin{itemize}
    \item \textbf{Adjustment of size}:Participants adjusted the size of the virtual cube through the manipulatives to match the actual cube they were holding.
\end{itemize}

\begin{figure}[t!]
\centering
\includegraphics[width=1.0\linewidth]{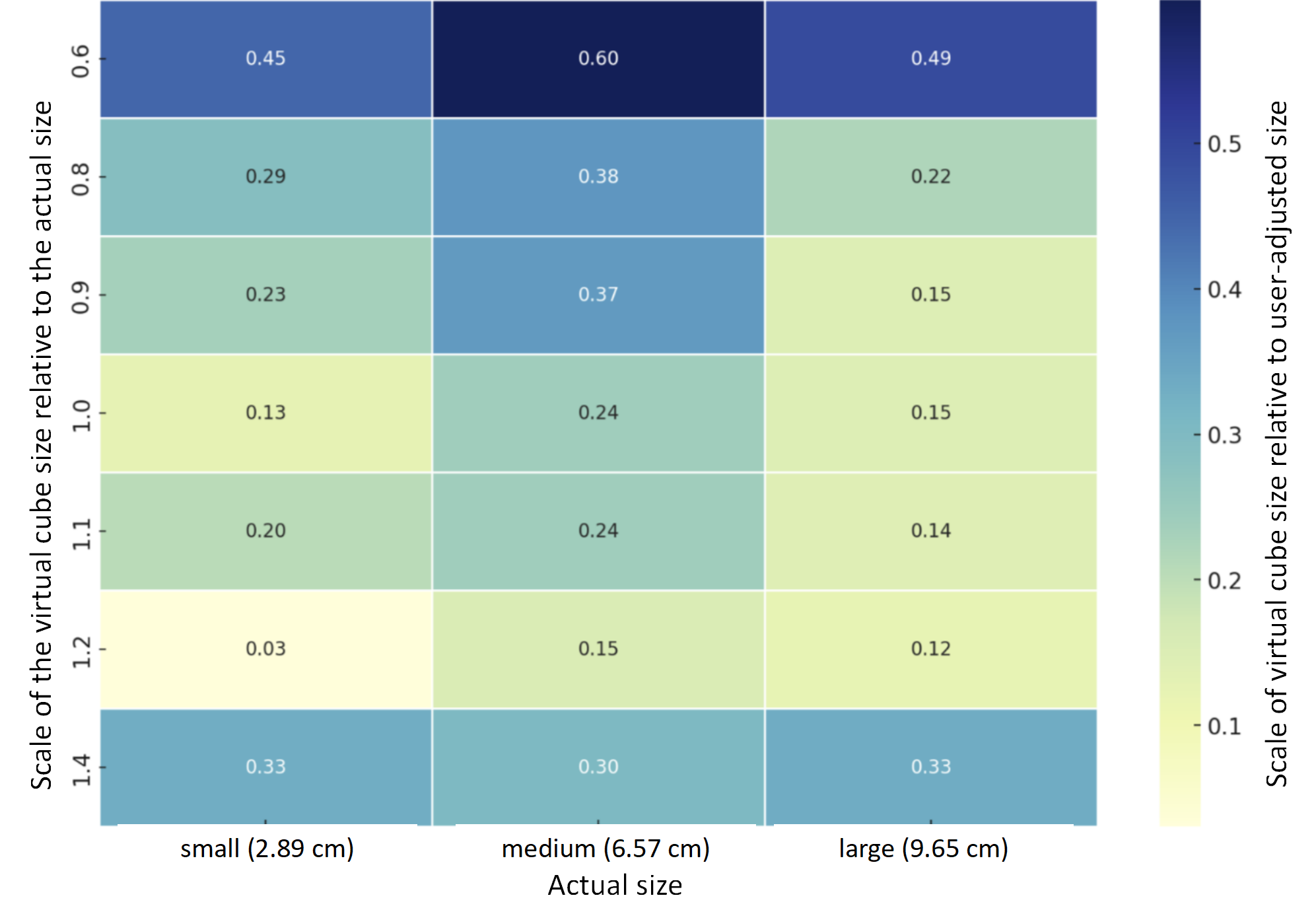}
\caption{Heat map of participant adjustment of size by participants. The X-axis represents the size of the cube touched by the participant in the real world (actual size), and the Y-axis represents the scale of the virtual cube size (virtual size) relative to the actual size in the VR environment (ranging from 60\% to 140\%). The color intensity of the heatmap indicates the proportion of adjustment made by the participant relative to the actual size when there is a perceived discrepancy between the actual and virtual sizes. Darker colors indicate greater adjustments. When the virtual size closely matches the actual size, the proportion of participant adjustments is smaller, suggesting that real tactile feedback can guide virtual perception to a certain extent. However, different actual sizes do not seem to affect this perception.}
\vspace{-20pt}
\label{fig:heat}
\end{figure}

At each stage, participants would experience the following seven sizes of VR cubes: 60\%, 80\%, 90\%, 100\%, 110\%, 120\%, and 140\%.

\subsection{Research findings}

\subsubsection{Data Analysis}

By analyzing the experimental data, we discovered several important results regarding the confirmation and adjustment ratio for different sizes. In the case of small, medium, and large sizes, the ratio of confirmation (T) and adjustment (F) varied significantly. For small sizes, there was a 48.57\% probability of confirmation, indicating that in almost half of the cases, participants believed the size of the virtual cube matched the actual cube they were holding, with no further adjustment needed. For medium sizes, the probability of confirmation was 45.71\%, showing a similar trend to the small size, where participants also believed the virtual cube was the correct size in about half of the cases. However, for large sizes, the probability of confirmation dropped to 25.71\%, suggesting that participants were less likely to confirm the size of the virtual cube matched the actual cube, requiring more frequent adjustments.

A detailed analysis reveals that participant confirmation probabilities are similar for small and medium sizes, suggesting a lower likelihood of participants breaking the illusion in these size ranges. The significantly lower confirmation probability for large sizes may be due to the more pronounced perceptual differences between the virtual and real environments for larger objects, leading to a greater need for adjustments.

The heat map in Figure \ref{fig:heat} illustrates the average degree of participant adjustment (Z value) for each combination of the size of the cube touched by the participant in the real world (X value) and scale of the virtual cube size (virtual size) relative to the actual size(Y value). Darker colors indicate a greater degree of adjustment, while lighter colors show a lesser degree of adjustment. The degree of participant adjustment is highest when the actual touch size is large (large X value) and the size seen in VR is small (small Y value). This suggests that participants are more likely to perceive inconsistencies between tactile and visual inputs in these scenarios. When the actual touch size and the size seen in VR are closer (similar X and Y values), participants adjust less, indicating a higher consistency between haptic and visual perceptions.

\begin{figure}[t!]
    \centering
    \includegraphics[width=1.0\linewidth]{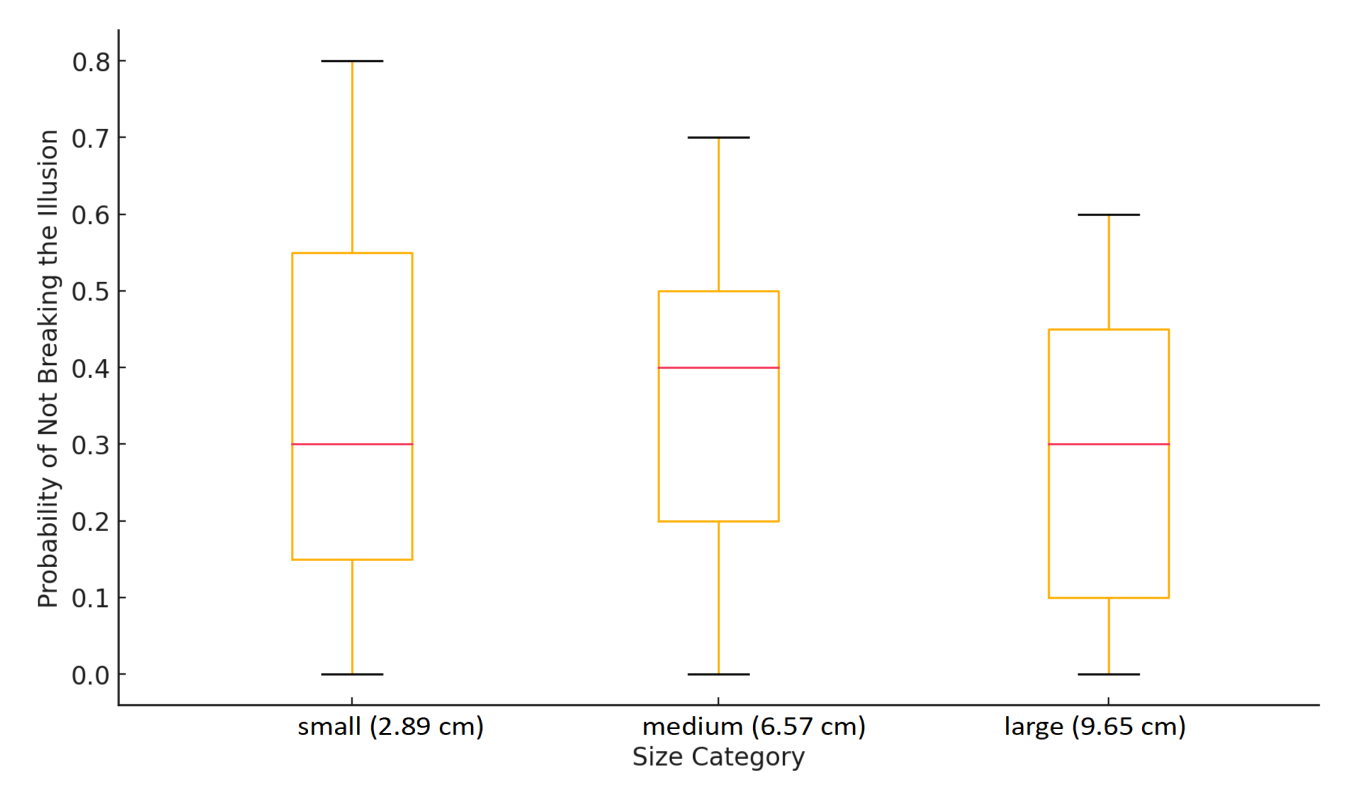}
    \caption{ Probability of not breaking the illusion by size category. The X-axis represents the actual size of the object touched by the participant, while the Y-axis represents the probability that participants do not perceive a difference between the virtual size and the actual size (i.e., the illusion is not broken) under different virtual sizes.}
    \label{fig:heat2}
\end{figure}

\begin{figure}[t!]
    \centering
    \includegraphics[width=1.0\linewidth]{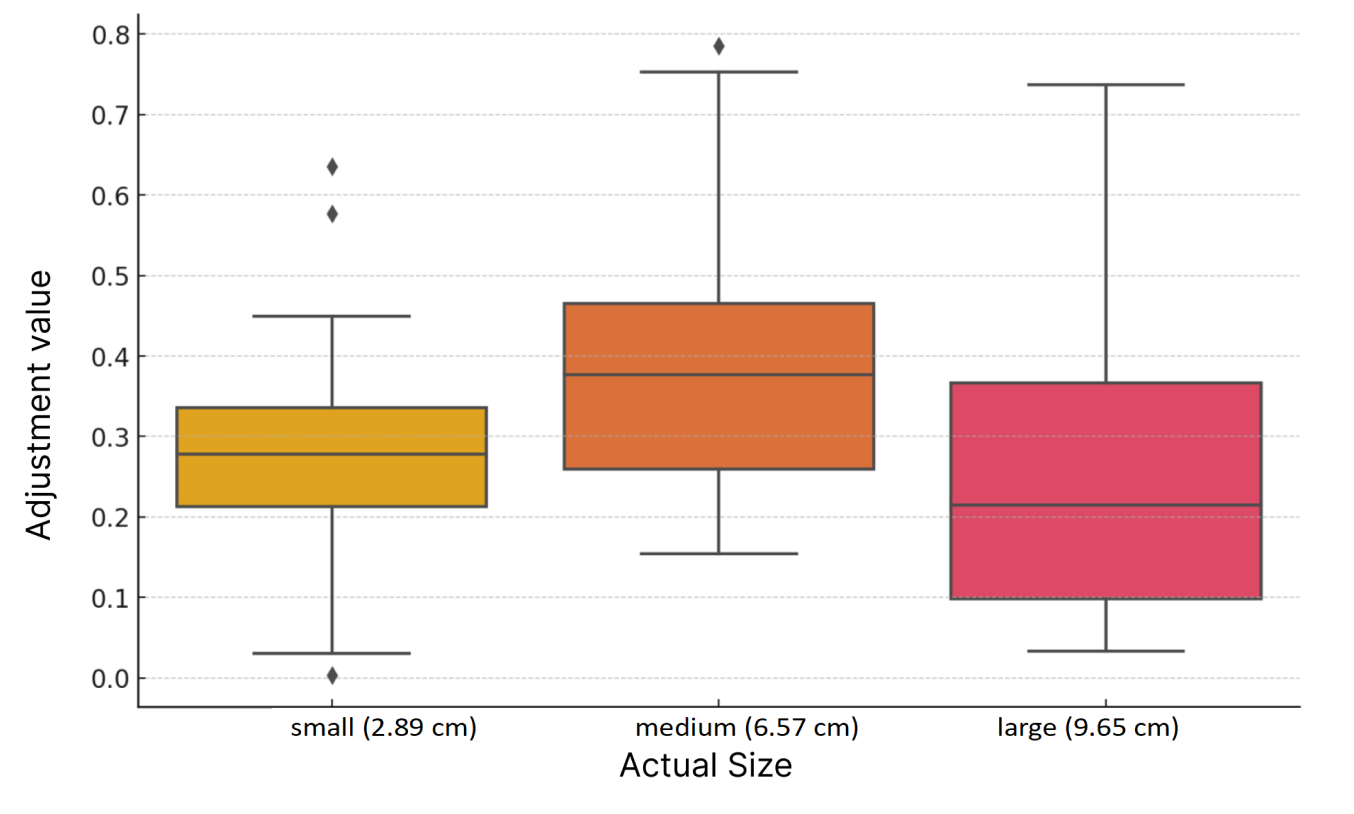}
    \vspace{-10pt} 
    \caption{Adjustment value by actual size. The X-axis represents the actual size of the object touched by the participant, while the Y-axis represents the adjustment value.}
    \vspace{-10pt} 
    \label{fig:heat22}
\end{figure}

Cardinality test results further elucidated these findings. The comparison between small and large sizes yielded a chi-square value of 2.998 with a p-value of 0.083, indicating that the difference in the probability of participant confirmation between these sizes is statistically close to significance but does not reach the 0.05 threshold. Comparing medium and large sizes, the chi-square value was 2.24 with a p-value of 0.134, showing that the difference in confirmation probability between these sizes is not statistically significant. From Figure \ref{fig:heat2}, we can see that the "Large" size category has the lowest probability of not breaking the illusion, with a slightly smaller range of probabilities compared to the "Medium" and "Small" categories. Therefore, the "Large" size condition is the one where the probability of not breaking the illusion is generally lower. However, it is important to note that our sample size was limited, and the results did not achieve statistical significance. Future studies with a larger sample size are needed to confirm these findings and provide more robust conclusions.

Then, we examined to what extent participants would adjust the size of the cubes in the case of breaking the illusion. We collected the value of the size adjusted by the participant and divided it by the value of the size of the cube presented in VR, which gave us the proportion of their adjustments (Adjustment value). Bigger adjustment value would indicate a bigger adjustment from participants.

Figure \ref{fig:heat22} shows a significant difference in the degree of adjustment between groups with different realistic sizes of touched cubes (P=0.033). The figure shows that for smaller sizes, the overall adjustments made by participants are the smallest, and the differences between different participants are also minimal, but there are more outliers.

\begin{figure}[t!]
    \centering
    \includegraphics[width=1.0\linewidth]{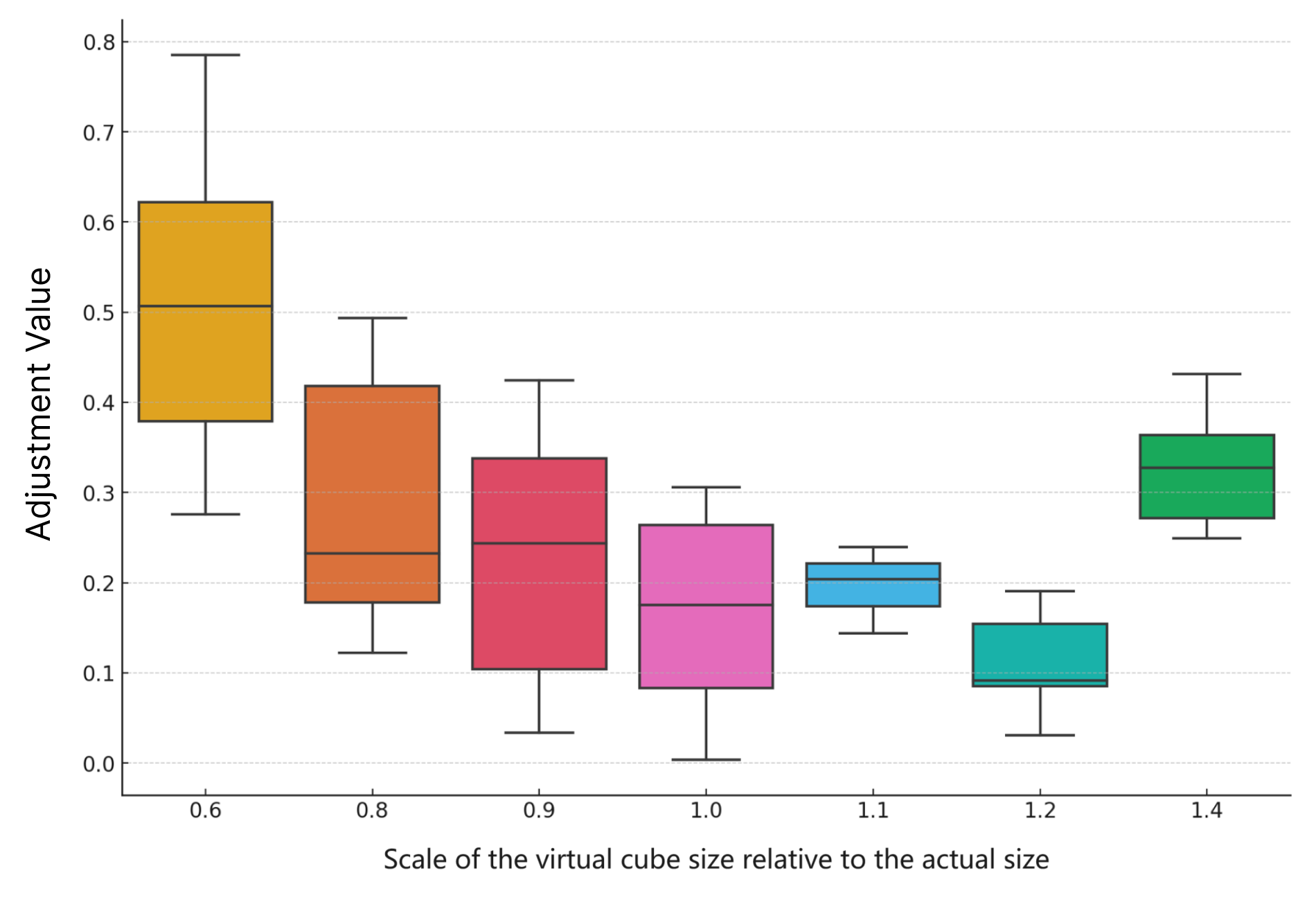}
    \vspace{-10pt} 
    \caption{Adjustment value by perceived size in VR. The X-axis represents the size of the virtual object touched by the participant, while the Y-axis represents the adjustment value.}
    \vspace{-10pt} 
    \label{fig:heat222}
\end{figure}

As illustrated in Figure \ref{fig:heat222}, the different sizes of the cube seen in VR have a significant effect on the extent to which participants resize the square in the VR environment (P<<0.05). When the cube size in VR is rendered between 110\% and 120\%, the adjustment values are smaller and more consistent. This indicates that although the illusion was broken, participants did not perceive a significant difference when changes in VR is small.

\subsubsection{Conclusions}

Our findings indicate that size-perception consistency is higher for small and medium sizes, with the virtual cube being perceived as the correct size in nearly half of the cases. Conversely, the likelihood of participants breaking the illusion is higher for large sizes, suggesting that the perceived difference between virtual and actual cube sizes is more pronounced in these cases, necessitating more adjustments. Although the difference in confirmation probability between small and large sizes approached statistical significance, the difference between medium and large sizes was not statistically significant. These results suggest that further validation and exploration can be achieved in future studies by increasing the sample size or improving the experimental design.

\section{Discussion and Limitation}

In this study, we investigated users' tolerance to tactile size distortions in VR environments. Previous research has established the dominance of visual input in tactile perception\cite{lederman2009haptic}, but the integrity of tactile perception in VR settings remains underexplored. Therefore, we developed a method to assess users' tolerance to size distortions by comparing various visual and tactile cues.

\subsection{Discussion}
This study investigates users' tolerance of discrepancy between the size of a model in VR and its size in reality, and how different sizes in reality affect this discrepancy. By analyzing the impact of various tactile and visual size combinations on participant adjustment behavior, we aim to gain deeper insights into user perception and behavior in VR environments.

We conducted multiple experiments, presenting virtual objects of various sizes in different visual contexts, accompanied by corresponding tactile feedback. The results indicate that when visual and tactile cues are synchronized, participants can tolerate significant size distortions. Our findings reveal that participants tend to overestimate the size of objects compared to their actual dimensions. This finding aligns with Norman et al. \cite{norman2004visual}, who proposed the theory of visual dominance, suggesting that visual information often takes precedence in multimodal perception conflicts. Additionally, we found that in VR environments, participants show a higher tolerance for overlapping objects compared to suspended objects. This might be because overlapping objects provide more visual cues, enhancing the effect of visual dominance\cite{ernst2002humans}.

These findings not only verify the dominant role of visual cues in tactile perception but also complement existing literature. For instance, \cite{ernst2002humans} demonstrated that when there is a conflict between visual and tactile information, visual information typically dominates. Our results extend this conclusion by showing that even in virtual reality environments, vision remains a critical factor influencing tactile perception.
We can compare these findings with existing studies to explore how they supplement or contradict current theories. The importance of tactile and visual consistency in immersive experiences has been highlighted by Slater et al.\cite{slater1993representations}. Our study further quantifies this consistency by examining participant adjustment behavior when there is a size discrepancy between what is touched and what is seen. This quantification provides a concrete measure of the impact of size consistency on user experience, supplementing the theoretical framework proposed by\cite{slater1993representations}.

Geiger et al. \cite{geiger} investigated the perception of objects of different sizes in VR, finding that larger objects are more likely to cause perceptual inconsistencies. Our results align with this finding, showing that when the tactile size is large, participants are more likely to perceive discrepancies and make significant adjustments. This correlation strengthens the argument that the size of objects plays a critical role in the perceived consistency of VR experiences.
However, our findings also reveal some contradictions with existing research. Bianchi-Berthouze et al. \cite{bianchi2007does} found that users have a higher tolerance for discrepancies with medium-sized objects in VR. In contrast, our study shows that even medium-sized objects can cause significant adjustments when the visual size is significantly smaller than the tactile size. This discrepancy suggests that participant tolerance for size inconsistencies may vary depending on specific experimental conditions, such as the type of task or individual differences among participants. Future research could explore these variables further to resolve these contradictions.
Our findings demonstrate that participants make varying degrees of adjustments based on the combination of tactile and visual sizes, highlighting the importance of maintaining consistency between what participants feel and see in VR. The quantified adjustments provide concrete evidence of how discrepancies affect VR users' perception, contributing to the existing body of research on immersive VR experiences.

By identifying the specific conditions under which users perceive the greatest discrepancies, our study offers practical insights for designing VR environments. Ensuring that tactile and visual sizes are consistent, especially for larger objects, can enhance user experience by reducing the need for adjustments and improving the perceived realism of VR. Moreover, these results have significant practical implications for VR design and applications, particularly in scenarios requiring precise tactile feedback. For example, understanding users' tolerance to size distortions can help design more realistic and effective virtual environments in scenarios such as surgical simulations or virtual product manipulations\cite{rosen2011surgical}.

\subsection{Limitations}
Despite providing valuable insights, this study has several limitations. First, we only considered square-shaped objects. Future research should include objects of different shapes, surface materials, and varying distances from the participant to assess the impact on VR environment tolerance. Second, we did not account for participants’ prior VR experience, hand size, or dominant hand, which could influence the results. Lastly, our sample size was limited. Future studies should involve a larger and more diverse sample to enhance the generalizability of the findings.

Additionally, due to the convenience of setting up the experimental environment and the ease of right-handed operation, we positioned the virtual cube slightly to the right in front of the participant. If the cube were centered or placed to the left, it might affect the experimental results. Future research should test how object placement at different positions relative to the participant affects their perception. Furthermore, the size of the virtual hand and the discrepancy between the virtual hand and the participant's actual hand size could also influence the outcomes, which should be considered in future studies.

Addressing these limitations can guide future research and contribute to improving VR interaction design and participant experience. Further exploration into the influence of different object properties and participant characteristics will provide a more comprehensive understanding of tactile perception in VR environments.

\section{Conclusion}
In this study, we investigated participants' tolerance to tactile size distortions in VR environments. Our findings indicate that when visual and tactile cues are synchronized, participants can tolerate significant size distortions, often overestimating the actual dimensions of objects. These results further validate the dominant role of visual cues in tactile perception and extend this understanding to virtual reality settings.

Our research contributes to the existing body of knowledge by demonstrating that visual dominance persists even with artificial tactile feedback in VR. This has significant implications for the design and application of VR systems, especially in scenarios requiring precise tactile feedback, such as surgical simulations and virtual product manipulations.

However, this study has several limitations, including the exclusive focus on square-shaped objects and the lack of consideration for participants' prior VR experience, hand size, and dominant hand. Future research should address these limitations by including a broader range of object shapes, surface materials, and participant characteristics. Additionally, larger and more diverse samples should be used to enhance the generalizability of the findings.

In conclusion, understanding participants' tolerance to tactile size distortions in VR environments can inform the design of more realistic and effective virtual experiences. Our study provides valuable insights that can guide future research and development in VR interaction design, not only improving the simulation of sandbox tools but also extending to other applications such as educational training, remote collaboration, and virtual reality gaming. Ultimately, this will enhance participant experience and broaden the applicability of VR technology across various domains.

\section{Acknowledgments}
This work is partially supported by Guangzhou-HKUST(GZ) Joint Funding Project (No.: 2024A03J0617), HKUST Practice Research with project title "RBM talent cultivation Exploration" (No.: HKUST(GZ)-ROP2023030) and HKUST(GZ) Metaverse Joint Innovation Laboratory.

\bibliographystyle{ACM-Reference-Format}
\bibliography{sample-base}










\end{document}